\documentclass[two column, prA]{revtex4}%
\usepackage{amsmath}
\usepackage{graphicx}
\usepackage{amsfonts}
\usepackage{amssymb}
\usepackage{color}%
\providecommand{\U}[1]{\protect\rule{.1in}{.1in}}
\ifx\pdfoutput\relax\let\pdfoutput=\undefined\fi
\newcount\msipdfoutput
\ifx\pdfoutput\undefined\else
\ifcase\pdfoutput\else
\ifx\paperwidth\undefined\else
\ifdim\paperheight=0pt\relax\else\pdfpageheight\paperheight\fi
\ifdim\paperwidth=0pt\relax\else\pdfpagewidth\paperwidth\fi
\fi\fi\fi
\begin{document}
\title{Unification of versions of photon quantum mechanics through Clifford spacetime algebra}
\author{Margaret Hawton}
\affiliation{Department of Physics, Lakehead University, Thunder Bay, ON, Canada, P7B 5E1}
\email{margaret.hawton@lakeheadu.ca}

\begin{abstract}
The Clifford spacetime algebraic description of Maxwell's equations is
reviewed and shown to give a unified picture of recently published versions of
photon quantum mechanics. Photon wave equations and a conserved four-current
are derived from the complexified standard Lagrangian. The equations of motion
and scalar product are found to be in good agreement with those obtained from
Fourier transformation of momentum space wave function and scalar product
[Phys. Rev. A \textbf{102}, 042201 (2020)].

\end{abstract}
\maketitle

\section{Introduction}

Photons are important particles in applications and tests of the foundations
of quantum mechanics. The Reimann-Silberstein (RS) vector is a popular choice
for the first quantized description of single photon states \cite{BB96} and
entangled pairs. We will show here that in $Cl_{1,3}$ space-time algebra the
RS vector reflects the geometry of space-time. The Clifford algebraic
description of electromagnetism is concise, but the primary advantage of this
approach that will be exploited here is to make a distinction between the
trivector, $i$, and the unit imaginary, $j$. This makes it possible to
separate the RS vector into its electric and magnetic field parts,
$\mathbf{F}=\mathbf{E}+ic\mathbf{B}$, while reserving $j$ for expressions such
as the circularly polarized plane wave $\left(  \mathbf{e}_{1}+j\lambda
e_{2}\right)  \exp\left[  -j\left(  \omega_{k}t-\mathbf{k\cdot x}\right)
\right]  $ where in this example $\mathbf{k}=k\mathbf{e}_{3}$, $\lambda=\pm1$
and $\epsilon=+$.

In Section II the $Cl_{1,3}$ space-time algebraic description of
electromagnetism will be reviewed and the equations of motion and scalar
product for relativistic photon quantum mechanics will be derived from the
complexified standard Lagrangian. This is complimenary to the derivation by
Bialynicki-Birula and Bialynicka-Birula \cite{BB2020} in which the
$\mathbf{k}$-space photon wave function and scalar product are Fourier
transformed to configuration space. In Section III we will examine the
relationship of space-time algebra to the recently published work on photon
quantum mechanics and compare the equations derived in Section II with the
results of \cite{BB2020}, In Section IV we will Conclude. SI units are used throughout.

\section{Photon quantum mechanics in Clifford spacetime algebra}

Paravectors and their reversals will be written as $U=U^{0}+\mathbf{U}$ and
$\overline{U}=U^{0}-\mathbf{U}$ respectively. The (noncommutative)
multiplication rule in $Cl_{1,3}$ Clifford algebra is%
\begin{equation}
UV=\left(  U^{0}V^{0}+\mathbf{U}\cdot\mathbf{V}\right)  +\ \left(
U^{0}\mathbf{V+U}V^{0}+i\mathbf{U}\times\mathbf{V}\right)  \label{Clifford}%
\end{equation}
where $U_{0}\mathbf{V+U}V_{0}$ and $\mathbf{U}\times\mathbf{V}$ are vectors
and $i\mathbf{U}\times\mathbf{V}$ is a bivector. Here $\mathbf{U}%
\cdot\mathbf{V}$ and $\mathbf{U}\times\mathbf{V}$ are the usual dot and cross
products. In an orthornormal 3D Cartesian basis $\left\{  \mathbf{e}%
_{1},\mathbf{e}_{2},\mathbf{e}_{3}\right\}  $ according to (\ref{Clifford})
the unit vectors satisfy $e_{i}e_{j}=\delta_{ij}+i\mathbf{e}_{i}%
\times\mathbf{e}_{j}$ so that $e_{i}e_{i}=1$ and $e_{i}e_{j}=-e_{j}e_{i}$ for
$i\neq j$. The trivector is $i=\mathbf{e}_{1}\mathbf{e}_{2}\mathbf{e}_{3}$.
Using these relations it is straightforward to verify that $i^{2}=-1$ and
$i\mathbf{e}_{3}=\mathbf{e}_{1}\mathbf{e}_{2}\mathbf{e}_{3}\mathbf{e}%
_{3}=\mathbf{e}_{1}\mathbf{e}_{2}$ is a bivector \cite{Baylis,Doran}. This
last relation can be permuted in cyclic order. If $U^{0}$ is a scalar and
$\mathbf{U}$ is a vector, in relativistic notation the paravector $U$ is
equivalent to the four-vector $U=\left(  U^{0},\mathbf{U}\right)  $. In tensor
notation $U=U^{\mu}$ is a contravariant four-vector, $\overline{U}=U_{\mu
}=g_{\mu\nu}U^{\nu}$ is the corresponding covariant four-vector and $g_{\mu
\nu}=g^{\mu\nu}$ is a $4\times4$ matrix with diagonal $\left(
1,-1,-1,-1\right)  $. Repeated Greek indices are summed over so, for example,
$U\overline{V}=U^{\mu}V_{\mu}=U^{0}V_{0}-\mathbf{U\cdot V}$ is an invariant scalar.

Writing the space-time coordinates, four-gradiant and four-potential
$x=\left(  ct,\mathbf{x}\right)  $, $\partial=\left(  \partial_{ct}%
,-\mathbf{\nabla}\right)  $ and $A=\left(  \phi/c,\mathbf{A}\right)  $ as the
paravectors $x=ct+\mathbf{x}$, $\partial=\partial_{ct}-\mathbf{\nabla}$ and
$A=\phi/c+\mathbf{A\ }$respectively, (\ref{Clifford}) gives%
\begin{align}
c\partial\overline{A}  &  =c\Lambda+\mathbf{E}+ic\mathbf{B}\label{divA}\\
\text{where }\Lambda &  =c^{-2}\partial_{t}\phi+\mathbf{\nabla}\cdot
\mathbf{A},\label{Lambda}\\
\mathbf{E}  &  =-\mathbf{\nabla}\phi-\partial_{t}\mathbf{A},\ \mathbf{B}%
=\mathbf{\nabla}\times\mathbf{A,} \label{EandB}%
\end{align}
$c$ is the speed of light in vacuum, $\partial_{ct}=\frac{1}{c}\frac{\partial
}{\partial t}$, $\Lambda$ determines the gauge, $\mathbf{E}$ and $\mathbf{B}$
are vectors and $ic\mathbf{B}$ is a bivector. The paravector (\ref{divA}) and
the RS vector $\mathbf{F}$ will be written as%
\begin{align}
F  &  =c\partial\overline{A}=c\Lambda+\mathbf{F},\label{F}\\
\mathbf{F}  &  =\mathbf{E}+ic\mathbf{B}. \label{Weber}%
\end{align}
Note especially that $i$ in this expression is a space-time trivector not the
unit imaginary. This separation of $F$ into its scalar, vector, bivector and
trivector parts is Lorentz reference frame dependent \cite{Doran}.

The wave equation and scalar product should be derivable from a Lagrangian. To
pass from a classical to a quantum mechanical interpretation of Maxwell's
equations and allow for the positive frequency wave functions in the
literature it will be assumed that the four-potential, $A$, and and its
complex conjugate, $A^{\ast}$, are independent. This is equivalent to treating
the real and imaginary parts of $A$ as independent. The standard, Fermi,
covariant photon and matter-photon interaction Lagrangian densities are then
\cite{CT}%
\begin{align}
\mathcal{L}_{std} &  =-\epsilon_{0}\left(  \mathbf{E}^{\ast}\mathbf{\cdot
E}-c^{2}\mathbf{B}^{\ast}\mathbf{\cdot B}\right)  =-\epsilon_{0}%
\mathcal{F}_{\mu\nu}^{\ast}\mathcal{F}^{\mu\nu},\label{Lstd}\\
\mathcal{L}_{Fermi} &  =\mathcal{L}_{std}-\epsilon_{0}c^{2}\Lambda
\Lambda^{\ast},\label{Lfermi}\\
\mathcal{L}_{cov} &  =-\epsilon_{0}c^{2}\left(  \partial^{\mu}A_{\nu}^{\ast
}\right)  \left(  \partial_{\mu}A^{\nu}\right)  ,\label{Lcov}\\
\mathcal{L}_{int} &  =-J_{m}^{\nu\ast}A_{\nu}-J_{m}^{\nu}A_{\nu}^{\ast
}\label{Lint}%
\end{align}
where $\epsilon_{0}$ is the dielectric permittivity,
\begin{equation}
\mathcal{F^{\mu\nu}=}\partial^{\mu}A^{\nu}-\partial^{\nu}A^{\mu}%
\label{Faraday}%
\end{equation}
is the antisymmetric Faraday tensor in which $\mathcal{F}^{00}=\mathcal{F}%
^{ii}=0$, $\mathcal{F}^{i0}=-\mathcal{F}^{0i}=\epsilon_{0}E_{i}$ and
$\mathcal{F}^{ij}=-\mathcal{F}^{ji}=\epsilon_{ijk}cB_{k}$ and $\epsilon_{ijk}$
is the Levi-Civita symbol. The Fermi Lagrangian density (\ref{Lfermi}) and the
covariant density (\ref{Lcov}) differ only by the four-divergence
$-\epsilon_{0}c^{2}\partial_{\mu}\left(  A^{\mu}\partial_{\nu}A^{\nu\ast
}-A^{\nu\ast}\partial_{\nu}A^{\mu}\right)  $ so they give identical equations
of motion. The matter current density $J_{m}$ is included for generality and
to allow discussion of emission of a photon by an atom and some recent papers
on the dressed photon wave function. Variation of the action leads to the
Lagrange equation of motion
\begin{equation}
\partial_{\mu}\frac{\partial\mathcal{L}}{\partial\left(  \partial_{\mu}A_{\nu
}^{\ast}\right)  }-\frac{\partial\mathcal{L}}{\partial A_{\nu}^{\ast}%
}=0\label{Leqn}%
\end{equation}
and its complex conjugate for any $\mathcal{L}\left(  A_{\nu},A_{\nu}^{\ast
},\partial^{\mu}A_{\nu},\partial^{\mu}A_{\nu}^{\ast}\right)  $. With the
conjugate to $A_{\nu}^{\ast}$ defined as%
\begin{align}
\Pi^{\mu\nu} &  =\partial\mathcal{L}/\partial\left(  \partial_{\mu}A_{\nu
}^{\ast}\right)  \text{ where}\label{PI}\\
\Pi_{std}^{\mu\nu} &  \equiv\frac{\partial\mathcal{L}_{std}}{\partial\left(
\partial_{\mu}A_{\nu}^{\ast}\right)  }=\epsilon_{0}\mathcal{F}^{\mu\nu
},\nonumber\\
\Pi_{cov}^{\mu\nu} &  \equiv\frac{\partial\mathcal{L}_{cov}}{\partial\left(
\partial_{\mu}A_{\nu}^{\ast}\right)  }=-\epsilon_{0}\partial^{\mu}A^{\nu
},\nonumber
\end{align}
the Lagrange equation (\ref{Leqn}) becomes $\partial_{\mu}\Pi^{\mu\nu
}=\partial\mathcal{L}/\partial A_{\nu}^{\ast}$. The global phase change
$A\longrightarrow e^{j\widehat{\epsilon}\alpha}A$ and $A^{\ast}\longrightarrow
e^{-j\widehat{\epsilon}\alpha}A^{\ast}$ is a symmetry of $\mathcal{L}_{std},$
$\mathcal{L}_{Fermi}$ and $\mathcal{L}_{cov}$. Since $\delta A\simeq
j\widehat{\epsilon}\alpha$ and $\delta A^{\ast}\simeq-j\widehat{\epsilon
}\alpha$, for infinitesimal $\alpha$ the Noether current $\Pi\delta A$
generated by this phase change is $J^{\mu}\propto j\Pi^{\mu\nu}%
\widehat{\epsilon}A_{\nu}^{\ast}-$ $j\Pi^{\mu\nu\ast}\widehat{\epsilon}A_{\nu
}$. This four-current will be discussed later in this Section. The operator
$\widehat{\epsilon}=j\left(  -\nabla^{2}\right)  ^{-1/2}\partial_{ct}$ and the
positive definite number density $J^{0}$ were introduced in
\cite{MostafazadehZamani,BabaeiMostafazadeh}. This operator extracts the sign
of the frequency, $\epsilon=\pm$ from $A$ and plays the same role as the
Paul-Dirac matrix $\gamma^{0}=\left(
\begin{array}
[c]{cc}%
I & 0\\
0 & -I
\end{array}
\right)  $ where $I$ is the $2\times2$ unit matrix. In a basis of positive and
negative frequency waves $\widehat{\epsilon}$ can be replaced with its
eigenvalues, $\epsilon=\pm$.

The Lagrange equation (\ref{Leqn}) with $\mathcal{L=L}_{std}+\mathcal{L}%
_{int}$ gives the wave equation
\begin{equation}
c\square\overline{A}=\overline{\partial}\left(  c\Lambda+\mathbf{F}\right)
=c\overline{\partial}\Lambda+Z_{0}\overline{J}_{m} \label{Wave}%
\end{equation}
where $\mu_{0}=\left(  \epsilon_{0}c^{2}\right)  ^{-1}$ is the magnetic
permeability, $Z_{0}=\sqrt{\mu_{0}/\epsilon_{0}}$ is the impedance of vacuum
and $\square=\overline{\partial}\partial=\partial\overline{\partial}%
=\partial_{ct}^{2}-\mathbf{\nabla}^{2}$. In the Coulomb gauge $\nabla
\cdot\mathbf{A}=0$, so there are no longitudinal modes and the scalar field
satisfying $\nabla^{2}\phi=-\rho/\epsilon_{0}$ responds instantaneously to
changes in charge density. Only the transverses modes $\lambda=\pm1$ propagate
at the speed of light and are second quantized in QED to allow creation and
annihilation of physical photons. Use of the covariant Lagrangian
$\mathcal{L}_{cov}$ or insertion of the Lorenz gauge condition $\Lambda=0$
into (\ref{Wave}) gives $\square A=\mu_{0}J_{m}$. In the Lorenz gauge all four
components of $A$ describing scalar, longitudinal and transverse $\lambda
=\pm1$ photon modes propagate at the speed of light and are second quantized
in QED.

The equation of motion for $A$ is gauge dependent but the $c\overline
{\partial}\Lambda$ terms on the left and right sides of (\ref{Wave}) cancel to
give a gauge independent equation for the RS vector,
\begin{equation}
\overline{\partial}\mathbf{F}=\mathbf{\nabla}.\mathbf{F}+\partial
_{ct}\mathbf{F}+i\mathbf{\nabla}\times\mathbf{F}=Z_{0}\overline{J}_{m}.
\label{divF}%
\end{equation}
When written as $\mathbf{F}=\mathbf{E}+ic\mathbf{B}$, (\ref{divF}) becomes
\begin{equation}
\mathbf{\nabla}.\mathbf{E}+ic\mathbf{\nabla}.\mathbf{B}+\partial
_{ct}\mathbf{E}-c\mathbf{\nabla}\times\mathbf{B}+i\partial_{t}\mathbf{B}%
+i\mathbf{\nabla}\times\mathbf{E}=Z_{0}\overline{J}_{m}. \label{ME1}%
\end{equation}
Since the scalar, trivector, vector and bivector terms are independent
(\ref{divF}) is equivalent to the Maxwell equations
\begin{align}
\mathbf{\nabla}\cdot\mathbf{B}  &  =0,\ \ \partial_{t}\mathbf{B}%
+\mathbf{\nabla}\times\mathbf{E}=0,\nonumber\\
\mathbf{\nabla}\cdot\mathbf{E}  &  =\frac{1}{\epsilon_{0}}\rho_{m}%
,\ \ \partial_{t}\mathbf{E}-c^{2}\mathbf{\nabla}\times\mathbf{B}=\frac
{1}{\epsilon_{0}}\mathbf{J}_{m}\mathbf{.} \label{MEs}%
\end{align}
The fields are in general complex and the complex conjugates of Eqs.
(\ref{Leqn}) to (\ref{MEs}) are also valid.

A basis of positive and negative frequency waves, $\epsilon=\pm$, and both
helicities, $\lambda=\pm1$, will be used. This allows for the possibility that
the wave function is real or restricted to positive frequencies. In the first
quantized theory described here $\epsilon=-$ represents a negative energy
photon. After second quantization a negative energy photon is replaced with
its positive energy antiphoton and annihilation of an $\epsilon=-$ photon with
wave vector $\mathbf{k}$ and helicity $\lambda$ appears as creation of an
antiphoton with wave vector $-\mathbf{k}$ and helicity $-\lambda$. To avoid
confusion, $\epsilon$ will be referred to as the sign of the frequency rather
than the sign of energy. Photons and antiphotons are indistinguishable and
$\epsilon$ and $\lambda$ are Lorentz invariants.

In vacuum the electric field describing a plane wave with helicity $\lambda$
and frequency $\epsilon\omega_{k}$ will be written as%
\begin{equation}
\mathbf{E}_{\mathbf{k}\lambda}^{\epsilon}\left(  t,\mathbf{x}\right)
=E\mathbf{e}_{\lambda}\left(  \widehat{\mathbf{k}}\right)  e^{-j\epsilon
\omega_{t}t+j\mathbf{k\cdot x}} \label{E}%
\end{equation}
where%
\begin{equation}
\mathbf{e}_{\lambda}\left(  \widehat{\mathbf{k}}\right)  =\frac{1}{\sqrt{2}%
}\left(  \mathbf{e}_{\theta}+j\lambda\mathbf{e}_{\phi}\right)  \text{,}
\label{e}%
\end{equation}
$\omega_{k}=ck$ with $k=\left\vert \mathbf{k}\right\vert $ and the
$\mathbf{k}$-space unit vectors $\left\{  \mathbf{e}_{\theta},\mathbf{e}%
_{\phi},\mathbf{e}_{\mathbf{k}}\right\}  $ form an orthonormal triad. It can
be verified by expansion that the real and imaginary parts of this field
rotate about the $\mathbf{k}$ direction, clockwise if $\epsilon\lambda=+1$ and
counterclockwise if $\epsilon\lambda=-1$. It follows from $\partial
_{t}\mathbf{B}+\mathbf{\nabla}\times\mathbf{E}=0$ that
\begin{equation}
c\mathbf{B}_{\mathbf{k}\lambda}^{\epsilon}=-j\epsilon\lambda\mathbf{E}%
_{\mathbf{k}\lambda}^{\epsilon} \label{B}%
\end{equation}
for any $\mathbf{k}$ and hence the RS vector is
\begin{equation}
\mathbf{F}_{\mathbf{k}\lambda}^{\epsilon}=\left(  1-ij\epsilon\lambda\right)
\mathbf{E}_{\mathbf{k}\lambda}^{\epsilon}. \label{PlaneWave}%
\end{equation}
Eqs. (\ref{E}) and (\ref{Weber}) then give $\mathbf{\nabla}\times
\mathbf{F}_{\mathbf{k}\lambda}^{\epsilon}=\lambda k\mathbf{F}_{\mathbf{k}%
\lambda}^{\epsilon}$ and $i\partial_{ct}\mathbf{F}_{\lambda\mathbf{k}%
}^{\epsilon}=-ij\epsilon k\left(  1-ij\epsilon\lambda\right)  \mathbf{E}%
_{\mathbf{k}\lambda}^{\epsilon}=\lambda k\mathbf{F}_{\mathbf{k}\lambda
}^{\epsilon}$ so individually and in linear combination the plane waves
(\ref{PlaneWave}) satisfy
\begin{equation}
i\partial_{ct}\mathbf{F}\left(  t,\mathbf{x}\right)  =\mathbf{\nabla}%
\times\mathbf{F}\left(  t,\mathbf{x}\right)  \text{,} \label{Feqn}%
\end{equation}
as they must for consistency with (\ref{divF}) in the absence of a matter
four-current. Thus any free space solution to (\ref{Feqn}) can be expanded in
the $\epsilon\lambda$-plane wave basis as the Fourier series
\begin{equation}
\mathbf{F}\left(  t,\mathbf{x}\right)  =\sum_{\epsilon,\lambda=\pm1}\int
d\mathbf{k}a_{\lambda}^{\epsilon}\left(  \mathbf{k}\right)  \mathbf{F}%
_{\mathbf{k}\lambda}^{\epsilon}\left(  t,\mathbf{x}\right)  .
\label{FinPlaneWaveBasis}%
\end{equation}
In the notation used here in which $i$ is a trivector and $j$ is the unit
imaginary, the single equation (\ref{Feqn}) describes positive and negative
frequency photons of both helicities. Eq. (\ref{FinPlaneWaveBasis}) allows us
to add the positive and negative and frequency waves describing the
indistinguishable photons and antiphotons to give real waves and take sums and
differences of circularly polarized waves to give linearly polarized waves.
When second quantized the expansion coefficient $a_{\lambda}^{+}\left(
\mathbf{k}\right)  $ in (\ref{FinPlaneWaveBasis}) is replaced with the
annihilation operator $\widehat{a}_{\lambda}\left(  \mathbf{k}\right)  $ and
$a_{-\lambda}^{-}\left(  -\mathbf{k}\right)  $ becomes the creation operator
$\widehat{a}_{\lambda}^{\dagger}\left(  \mathbf{k}\right)  $ that can act on
the vacuum state $\left\vert 0\right\rangle $ to give the Schr\"{o}dinger
picture (SP) one-photon plane wave state%
\begin{equation}
\left\vert 1_{\mathbf{k}\lambda}\right\rangle =\widehat{a}_{\lambda}^{\dagger
}\left(  \mathbf{k}\right)  \left\vert 0\right\rangle . \label{Creation}%
\end{equation}

The Noether current generated by the phase change $A^{\epsilon}\longrightarrow
e^{j\epsilon\alpha}A^{\epsilon}$ is $J^{\mu}\propto j\epsilon\Pi^{\mu\nu
}A_{\nu}^{\ast}-$ $j\epsilon\Pi^{\mu\nu\ast}A_{\nu}$ with $\Pi_{std}$ and
$\Pi_{cov}$ given by (\ref{PI}). The covariant Lagrangian density
$\mathcal{L}_{cov}$ gives the covariant photon four-current density
\begin{equation}
J_{cov}^{\mu}\left(  x\right)  =-\frac{j\epsilon_{0}c}{\hbar}\sum
_{\epsilon=\pm}\epsilon A_{\nu}^{\epsilon\ast}\left(  x\right)
\overleftrightarrow{\partial}^{\mu}A^{\epsilon\nu}\left(  x\right)
\label{Jcov}%
\end{equation}
in the $\epsilon=\pm$ basis. In this expression $f\overleftrightarrow{\partial
}^{\mu}g=f\left(  \partial^{\mu}g\right)  -\left(  \partial^{\mu}f\right)  g$.
In the Lorenz gauge the scalar and longitudinal contributions to $J_{cov}%
^{\mu}$ cancel, leaving only the contributions of the transverse modes
\cite{MaxwellQM}. If the standard Lagrangian density $\mathcal{L}_{std}$ in
the Coulomb gauge is used instead the Noether current is%
\begin{align}
J^{\mu}\left(  x\right)   &  =\frac{j\epsilon_{0}c}{\hbar}\sum_{\epsilon
,\lambda=\pm}\epsilon\left(  \mathbf{E}_{\lambda}^{\epsilon\ast}\left(
x\right)  \cdot\mathbf{A}_{\lambda}^{\epsilon}\left(  x\right)  ,\right.
\label{J}\\
&  \left.  -\mathbf{B}_{\lambda}^{\epsilon\ast}\left(  x\right)
\times\mathbf{A}_{\lambda}^{\epsilon}\left(  x\right)  +\mathbf{E}_{\lambda
}^{\epsilon\ast}\left(  x\right)  \phi_{\lambda}^{\epsilon}\left(  x\right)
\right)  +c.c.\nonumber
\end{align}
where $c.c.$ is the complex conjugate. Since $\mathbf{E}_{\lambda}^{\epsilon
}\left(  x\right)  =-\partial_{t}\mathbf{A}_{\lambda}^{\epsilon}\left(
x\right)  $ for the transverse modes $\lambda=\pm1$, $J^{0}\left(  x\right)
=J_{cov}^{0}\left(  x\right)  $. In a general gauge longitudinal modes are not
excluded and $J^{\mu}\propto j\sum_{\epsilon=\pm1}\epsilon\left(
\mathbf{E}^{\epsilon\ast}\cdot\mathbf{A}^{\epsilon},-c\mathbf{B}^{\epsilon
\ast}\times\mathbf{A}^{\epsilon}+\mathbf{E}^{\epsilon\ast}\phi^{\epsilon
}\right)  +c.c.$. The four-current density $J^{\mu}\left(  x\right)
\propto\mathcal{F}^{\mu\nu}\left(  x\right)  A_{\nu}\left(  x\right)  $ was
first obtained in \cite{HawtonMelde} where it was used to derive a Hermitian
number density operator. A continuity equation with a charged matter source
can be derived by evaluating $\partial_{\mu}J^{\mu}$ and substituting the
equation of motion (\ref{Wave}) to give%
\begin{equation}
\partial_{\mu}J^{\mu}\left(  x\right)  =-\frac{j\mu_{0}}{\hbar c}A_{\nu}%
^{\ast}\left(  x\right)  J_{m}+c.c.. \label{continuity}%
\end{equation}
In the absence of a charged matter source (\ref{continuity}) verifies that the
four-currents (\ref{Jcov}) and (\ref{J}) satisfy continuity equations and
photon number is conserved.

For vector potentials $A_{1}$ and $A_{2}$ with $\epsilon$ and $\lambda$
components $A_{1\lambda}^{\epsilon}$ and $A_{2\lambda^{\prime}}^{\epsilon
^{\prime}}$ the scalar product at a fixed time $t$, derived from the spatial
integral of the zeroth component of the four-current (\ref{J}) is
\cite{MaxwellQM}
\begin{equation}
\left(  A_{1},A_{2}\right)  _{t}=\frac{j2\epsilon_{0}}{\hbar}\sum
_{\epsilon,\lambda=\pm1}\epsilon\int_{t}d\mathbf{xA}_{1\lambda}^{\epsilon\ast
}\left(  x\right)  \cdot\mathbf{E}_{2\lambda^{\prime}}^{\epsilon^{\prime}%
}\left(  x\right)  \delta_{\epsilon\epsilon^{\prime}}\delta_{\lambda
\lambda^{\prime}}.\label{EA}%
\end{equation}
Substitution of $\mathbf{E}=-\partial_{t}\mathbf{A}$ and the one-photon
amplitude $E=\frac{1}{\left(  2\pi\right)  ^{3}}\sqrt{\frac{\hbar}%
{\epsilon_{0}}}$ in (\ref{E}) gives the one-photon SP four-potential
\begin{equation}
\mathbf{A}_{\lambda}^{\epsilon\mu}\left(  t,\mathbf{x}\right)  =j\sqrt
{\frac{\hbar}{\epsilon_{0}}}\int_{t}\frac{d\mathbf{k}}{\left(  2\pi\right)
^{3}2\omega_{k}}\mathbf{e}_{\lambda}\left(  \mathbf{k}\right)  a_{\lambda
}^{\epsilon}\left(  t,\mathbf{k}\right)  e^{-j\epsilon\omega_{t}%
t+j\mathbf{k\cdot x}}\label{A}%
\end{equation}
and the scalar product (\ref{EA}) as%
\begin{equation}
\left(  A_{1},A_{2}\right)  _{t}=\sum_{\epsilon,\lambda=\pm1}\int_{t}%
\frac{d\mathbf{k}}{\left(  2\pi\right)  ^{3}}a_{1\lambda}^{\epsilon\ast
}\left(  t,\mathbf{k}\right)  a_{2\lambda}^{\epsilon}\left(  t,\mathbf{k}%
\right)  .\label{EAk}%
\end{equation}

In addition to a Hilbert space, standard quantum mechanics requires operators
representing the relevant physical observables. The momentum, Hamiltonian,
angular momentum and Lorentz reference frame are symmetries of the Lagrangian
density (\ref{Lstd}) \cite{Weinberg,HawtonBaylis,WignerLittleGroup}. The
corresponding operators are the generators of translations in space,
translation in time, rotations and boosts that should satisfy the commutation
relations $\left[  \widehat{J}_{i},\widehat{J}_{j}\right]  =j\hbar
\epsilon_{ijk}\widehat{J}_{k},$ $\left[  \widehat{J}_{i},\widehat{K}%
_{j}\right]  =j\hbar\epsilon_{ijk}\widehat{K}_{k},$ $\left[  \widehat{K}%
_{i},\widehat{K}_{j}\right]  =-j\hbar\epsilon_{ijk}\widehat{J}_{k},$ $\left[
\widehat{J}_{i},\widehat{P}_{j}\right]  =j\hbar\epsilon_{ijk}\widehat{P}_{k},$
$\left[  \widehat{K}_{i},\widehat{P}_{j}\right]  =j\hbar\delta_{ij}%
\widehat{H},$ $\left[  \widehat{K}_{i},\widehat{H}\right]  =-j\hbar
\widehat{P}_{i},$ $\left[  \widehat{J}_{i},\widehat{H}\right]  =\left[
\widehat{P}_{i},\widehat{H}\right]  =\left[  \widehat{P}_{i},\widehat{P}%
_{j}\right]  =0$ for $i=1,2,3$ \cite{Weinberg}. In the SP these operators are
time independent. In configuration space $\widehat{\mathbf{p}}=-j\hbar
\mathbf{\nabla}$\textbf{, }$\widehat{p}=\hbar\sqrt{-\nabla^{2}}$,
$\widehat{\mathbf{J}}=j\hbar\left(  \mathbf{x}\times\nabla+\widehat{\mathbf{S}%
}\right)  $ as in \cite{CT} compliment B$_{\text{I}}$\ where
$\widehat{\mathbf{S}}$ is the spin operator. In momentum space
$\widehat{\mathbf{p}}=\hbar\mathbf{k}$, $\widehat{p}=\hbar k$,
$\widehat{\mathbf{J}}=-j\hbar\left(  \mathbf{k}\times\nabla_{\mathbf{k}%
}+\widehat{\mathbf{S}}\right)  $ and $\widehat{\mathbf{K}}=j\hbar\left(
k\nabla_{\mathbf{k}}+\mathbf{k}\times\widehat{\mathbf{S}}\right)  $. The
helicity operator is $\widehat{\lambda}=\widehat{p}^{-1}\widehat{\mathbf{S}%
}\cdot\widehat{\mathbf{p}}$. A Hamiltonian operator is also needed, but
(\ref{Feqn}) is not of Schr\"{o}dinger form required for description of
unitary time evolution since its left hand side is multiplied by the trivector
rather than the unit imaginary. This can be easily be remedied by multiplying
(\ref{Feqn}) by $-ji$. Using $\mathbf{a}\times\mathbf{b}=-j\left(
\mathbf{a}\cdot\mathbf{S}\right)  $ to write $\mathbf{\nabla}\times\mathbf{F}$
as $-j\left(  \widehat{\mathbf{S}}\cdot\mathbf{\nabla}\right)  \mathbf{F}$,
(\ref{Feqn}) multiplied by $-ji$ becomes $j\hbar\partial_{t}\mathbf{F}%
=\epsilon c\widehat{p}\mathbf{F}$ which is of the Schr\"{o}dinger form
$j\hbar\partial_{t}\mathbf{F}=\widehat{H}\mathbf{F}$ for the Hamiltonian
$\widehat{H}=\epsilon c\widehat{p}$. This Hamiltonian operator describes the
time development of both positive and negative frequency waves.

\section{Relationship to previous work on configuration space photon quantum
mechanics}

In this Section the BB-Sipe photon wave function as extended by Smith and
Raymer and a number of recent proposals for photon quantum mechanics are
discussed in approximate chronological order
\cite{BB96,Sipe,SmithRaymer,Keller,Vicino,HawtonPosOp,Saunders,MostafazadehZamani,DebierreDurt,HawtonDebierre,BabaeiMostafazadeh,Kiessling,Sebens,MaxwellQM}%
.This work was motivated by a desire to understand the relationship of photon
quantum mechanics as formulated in
\cite{HawtonDebierre,BabaeiMostafazadeh,MaxwellQM} to the Dirac-like
formulation in \cite{BB96,Vicino,Kiessling}, leading to a study of spacetime
algebra as presented in \cite{Baylis,Doran}, an appreciation of the
fundamental geometrical significance of the RS-vector, and the realization
that the trivector, $i$, need not be equated to the unit imaginary, $j$.

Motivated by the potential for applications in quantum optics,
Bialynicki-Birula and Sipe \cite{BB96,Sipe} independently proposed that the
positive energy part of the six-component RS vector normalized to the average
energy of a single photon is a good wave function "for practical purposes".
Smith and Raymer \ state that "Maxwell unknowingly discovered a correct
relativistic, quantum theory for the light quantum". They extended the BB-Sipe
formalism to include a biorthogonal scalar product and modes that form an
orthonormal set. Their scalar product is of the form (\ref{EA}). It is
proportional to the spatial integral of the field multiplied by its dual. In
$\mathbf{k}$-space the potential is reduced by a factor $k^{-1}$ relative to
the field so in configuration space the potential is nonlocal with a kernel
proportional to $\left\vert \mathbf{x}-\mathbf{x}^{\prime}\right\vert ^{-2}$.
This is a feature of any biorthogonal scalar product of the form (\ref{EA}).
The authors note that this provides a link to the photon counting operator
derived in \cite{HawtonMelde}.

Newton and Wigner solved the position eigenvector problem for spin $\frac
{1}{2}$ electrons and spin $0$ Klein-Gordon particles, but their method failed
in the case of spin $1$, that is for photons \cite{NewtonWigner}. Following
their method but with omission of their spherical symmetry assumption, a
photon position operator with commuting components, $\widehat{\mathbf{x}}$,
was derived in \cite{HawtonPosOp} and it was found in \cite{HawtonBaylis} that
the biorthogonal photon position eigenvectors are cylindrically symmetrical.
This cylindrical symmetry reflects the $e(2)$ symmetry of the photon little
group \cite{WignerLittleGroup}. To take this little group symmetry into
account, "invariant under rotations about the origin" in NW's assumptions
should be generalized to "invariant under symmetry operations of the Wigner
little group". For massive particles these operations are just those
considered by NW, while for photons the little group consists of rotations
about some conveniently chosen $3$-axis and two boosts with compensating
rotations \cite{Weinberg,WignerLittleGroup}. It was proved that $\left\{
\widehat{J}_{3},\widehat{x}_{1},\widehat{x}_{2}\right\}  $ is a realization of
the algebra. The photon position eigenvectors derived in \cite{HawtonPosOp}
satisfy these modified postulates and describe the physics of optical beams
\cite{HawtonBaylis,WignerLittleGroup} that have a corresponding axis of symmetry.

Mostafazadeh and collaborators \cite{MostafazadehZamani,BabaeiMostafazadeh}
derived a positive definite number density by defining the conjugate field
$A_{c}\equiv j\widehat{D}^{-1/2}\partial_{ct}A=\widehat{\epsilon}A$ with
${\widehat{D}\equiv-\nabla^{2}}${.} In the $\epsilon=\pm$ basis used here
$A_{c}^{\epsilon}=\widehat{\epsilon}A^{\epsilon}=\epsilon A^{\epsilon}$.{ }It
can be verified by substitution that if $A$ satisfies Maxwell's wave equation,
$A_{c}$ also satisfies this equation and that the photon four-current $J^{\mu
}\left(  x\right)  =-\frac{j\epsilon_{0}c}{\hbar}A_{\nu}^{\ast}\left(
x\right)  \overleftrightarrow{\partial}^{\mu}A_{c}^{\nu}\left(  x\right)  $
satisfies a continuity equation. The photon density $J^{0}\left(  x\right)  $
is positive definite. Based on this number density they define a general
scalar product for which all state vectors have a real positive norm.

Tamburini and Vicino \cite{Vicino} write (\ref{Feqn}) in covariant form using
the Pauli matrices that are a representation of the $Cl_{1,3}$ spacetime
algebra. They conclude that the photon wave function approach can lead only to
results described by standard QED. Standard QED will be obtained by second
quantization of the theory described here in Section II.

Refs. \cite{HawtonDebierre,BabaeiMostafazadeh,MaxwellQM} are based on the
second order wave equation for $A$ and a scalar product that can be written in
the form (\ref{EA}). They include a position operator and give an expression
for the probability density as a function of $\mathbf{x}$ on the
$t$-hyperplane. Babaei and Mostafazadeh derived Hamiltonian, position,
helicity, momentum and chirality operators. Their Hamiltonian in
\cite{BabaeiMostafazadeh} is equivalent to $\widehat{H}$ derived in Section II
and their position operator is derived in the Heisenberg Picture.

The Newton Wigner position eigenvectors are nonlocal in configuration space
and they are not Lorentz covariant. Hawton and Debierre
\cite{HawtonDebierre,MaxwellQM} proved that the NW factor $k^{1/2}$ is a
\ consequence of non-covariant normalization of the plane wave basis and that
it can be eliminated if invariant plane wave normalization is used
\cite{ItzyksonZuber}. The position eigenvectors and their duals are potentials
that transforms as Lorentz four-vectors and fields proportional to
$\mathbf{E}_{x^{\prime}\lambda}^{\epsilon}$ that at $t=0$ is a pulse of
twisted light localizable in an arbitrarily small region. They derive an
expression for a positive and negative frequency position probability
amplitudes. These functions have a nonlocal imaginary part instantaneously
masked by destructive interference at $t=0$ and that is nonzero everywhere for
$t\neq0$, consistent with the Hegerfeldt theorem \cite{Hegerfeldt}. True
rather than apparent localization is achieved by adding positive and negative
frequency waves to give a probability amplitude that is localized at $t=0$ and
propagates causally \cite{HawtonDebierre,MaxwellQM}.

Kiessling and Tahvildar-Zadeh \cite{Kiessling} write the photon wave equation
in first order Dirac form using gamma matrices and derive a rank-two bi-spinor
photon wave function, $\psi_{ph}$, from a Lagrangian. Their $\sigma\left(
\mathbf{\xi}\right)  $ and $\sigma\left(  a\right)  $ in $\psi_{ph}$ are Pauli
representations of the paravectors $\mathbf{\xi}$ and $a$. They identify
$\mathbf{e}+i\mathbf{b}$ in $\mathbf{\xi}_{+}=\mathbf{\xi}_{-}=\left(
0,\mathbf{e}+i\mathbf{b}\right)  ^{T}$ with the RS vector $\mathbf{F}%
=\mathbf{E}+ic\mathbf{B}$. Using (\ref{divA}) and writing $\xi_{\pm}$ as
$F_{\pm}$, their equation Eq. (4.5),
\begin{equation}
-i\hbar\gamma^{\mu}\partial_{\mu}\psi_{ph}=m_{\swarrow}\Pi\psi_{ph}
\label{Dirac}%
\end{equation}
with $\Pi$ defined as a projection operator onto the diagonal of $\psi_{ph}$,
can be written as%
\begin{equation}
\left(
\begin{array}
[c]{cc}%
\frac{m_{\swarrow}c}{\hbar}\left(  c\partial\overline{A}^{+}-F^{+}\right)  &
\partial\overline{F}^{-}\\
\overline{\partial}F^{+} & \frac{m_{\swarrow}c}{\hbar}\left(  c\overline
{\partial}A^{-}-\overline{F}^{-}\right)  _{-}%
\end{array}
\right)  =0 \label{KT}%
\end{equation}
where the factor $\frac{m_{\swarrow}c}{\hbar}$ ensures that all the elements
of (\ref{KT}) have the same dimensions. Then, according to (\ref{divA}) to
(\ref{MEs}), their $\mathbf{\xi}_{+}=F^{+}$ and $\mathbf{\xi}_{-}^{\ast
}=\overline{F}^{-}$ satisfy Maxwell's equations and are four-gradients of
$A^{\pm}$. Eq. (\ref{KT}) incorporates (\ref{divA}) and (\ref{divF}) of
Section II into a single first order matrix equation. Kiessling and
Tahvildar-Zadeh derive a non-negative Born-rule-type quantum probability based
on a four-current that transforms as a Lorentz four-vector by projecting a
second rank tensor onto a future-oriented timelike four-vector. However it is
interesting to note that their $j_{Z}^{0}$ for the case $G=jb\gamma_{5}$ is
proportional to the number density in (\ref{J}). Eq. (4.7) of \cite{Kiessling}%
, equivalent to (\ref{divF}) here, is its Pauli-Dirac representation. It is
not covariant under the full Lorentz group and their (4.9) is added to
supplement it. They conclude that the doubled-up self-dual $\mathbf{F}$ cannot
be a wave function in a complex Hilbert space. If $i$ were to be identified
with $j$ in (\ref{PlaneWave}), $\mathbf{F}=0$ in the case $\epsilon\lambda=-1$
so (\ref{F}) here would have to be doubled-up to give a six-component vector
describing photons of both helicities. The need for doubling is avoided in
Section II by making a distinction between $i$ and $j$ so that the single
$3$-vector equation (\ref{Feqn}) describes photons and antiphotons of both
helicities and $\mathbf{F}$ is a good wave function. Their equation
(\ref{Dirac}) makes explicit the relationship of photon quantum mechanics to
the Dirac theory of electrons and positrons.

Wharton gives an insightful discussion of the difficulties encountered when
imposing two initial/boundary conditions on the solutions to the KG equation
\cite{Wharton}. In his formalism boundary conditions are imposed at two
different times. The complex Maxwell wave equation for $A$ is also second
order and requires two boundary conditions. It is conventional in classical
electromagnetism to reject the advanced Green's function and keep only the
retarded solution but careful examination of the mathematics shows that the
advanced component determines the past history of the field, while the
retarded component determines its future history \cite{Schweber}. The device
of the retarded solution is illusory and boundary condition are required to
determines contributions from homogeneous solutions to the wave equation.
Understanding of the relationships amongst boundary conditions, causality and
emission by a localized source is still work in progress.

In the Landau-Peierls (LP) formalism the RS vector is divided by a factor
proportional to $\sqrt{k}$ to give a function whose absolute square has
dimensions of number density. This approach was found to be not very useful,
since the LP wave function is non-locally connected to the classical
electromagnetic field and the current density \cite{SmithRaymer}. Recently
Sebens \cite{Sebens} postulated a Dirac-like equation of this type and again
concluded that this approach is unsatisfactory because the probabilities
derived from it do not always transform properly under Lorentz
transformations. It appears that the LP function is not an acceptable photon
wave function.

The photon wave function approach has been applied to emission of a photon by
an atom and the propagation of electromagnetic waves in non-absorptive
continuous media. Bialynicki-Birula derived the RS vector in a medium and in
curved space \cite{BB96}. Sipe \cite{Sipe} and Debierre \cite{DebierreDurt}
calculate the photon wave function emitted by an atom in an excited state.
Their positive frequency wave functions are intrinsically nonlocal, and this
leads to apparent violations of Einstein causality. Keller provides an
extensive discussion of the localization problem with an emphasis on the near
field regime \cite{Keller}. In \cite{Monken} the dressed photon wave function
was second quantized and the quantum state of the photons generated in
parametric down conversion was found to\ be intuitive but in agreement with
previous QED based treatments. This approach was later applied to Raman
scattering in which a phonon created in a Stokes process is absorbed by a
second anti-Stokes photon to yield a highly correlated photon pair.

To sum up, all of the theories discussed this Section are consistent with the
spacetime formulation in Section II and together they provide a complete
picture of configuration space photon quantum mechanics. Bialynicki-Birula and
Sipe recognized the importance of the RS vector. Hawton and Baylis derived a
photon position operator with commuting components and investigated its
properties \cite{HawtonPosOp,HawtonBaylis}. Tamburini and Vicino derived
(\ref{Feqn}) from the standard Lagrangian density and came to the correct
conclusion that any first quantized theory of the photon could only lead to
QED. Mostafazadeh and co-workers extended the Hilbert space to negative
frequencies by deriving a number density that is positive definite for both
positive and negative frequency waves
\cite{MostafazadehZamani,BabaeiMostafazadeh}. Hawton and Debierre used
biorthogonality to derive covariant position eigenvectors and concluded that
only the real part of the positive and negative frequency position
eigenvectors are truly localized and propagate causally
\cite{HawtonDebierre,MaxwellQM}. Kiessling and Tahvildar-Zadeh derived a Dirac
type wave equation from a Lagrangian \cite{Kiessling}. Refs.
\cite{HawtonDebierre,BabaeiMostafazadeh,MaxwellQM} include a position
eigenvectors, while in \cite{BabaeiMostafazadeh,Kiessling} a complete theory
of quantum mechanics was emphasized.

Photon quantum mechanics is well established in $\mathbf{k}$-space.
Bialynicki-Birula and Bialynicka-Birula \cite{BB2020}\ Fourier transformed the
$\mathbf{k}$-space wave function and scalar product to configuration space to
give a wave function and scalar product that are in excellent agreement with
those obtained in Section II. The equation of motion (\ref{divF}) or
(\ref{Feqn}) with $\mathbf{\nabla}\cdot\mathbf{F}=0$ here is equivalent to (7)
in \cite{BB2020}. The Fourier expanded RS wave function
(\ref{FinPlaneWaveBasis}) reduces to (10) in \cite{BB2020} if the trivector
$i$ is equated to the unit imaginary $j$ in (\ref{FinPlaneWaveBasis}). In the
notation used here their $\mathbf{k}$-space scalar product (2) is the
biorthogonal form $\left\langle k^{-1}f_{1}|f_{2}\right\rangle $ and its
Fourier transformed configuration space equivalent (20) is also biorthogonal
and equivalent to (\ref{EA}) here.

\section{Conclusion}

In Section II equations of motion for $A^{\mu}$ and the RS vector
$\mathbf{F}=\mathbf{E}+ic\mathbf{B}$ and a continuity equation for a positive
definite photon number density were derived from the complexified standard
Lagrangian. The reverse of this derivation in which the $\mathbf{k}$-space
photon wave function and scalar product are transformed to configuration space
\cite{BB2020} compliments and supports the results obtained here. So what do
these recent formulations of photon quantum mechanics contribute to our
understanding of electromagnetism and QED? Their very existence implies that
the photon is an elementary particle like any other and hence the photon is a
good candidate for tests of the foundations of quantum mechanics. The photon
wave function has been found to be useful in quantum optics for practical
purposes, so perhaps this extension of the BB-Sipe wave function to full
photon quantum mechanics and to non-dispersive and inhomogeneous media will
find additional applications. These first quantized theories are consistent
with QED as required for agreement with all experiments performed to date. It
appears that a quantum interpretation of "classical" electromagnetic theory
does indeed exist, as hinted at in Chapter 2 of Photons and Atoms where
transverse angular momentum is written as the expectation value of
$\widehat{\mathbf{J}}$ \cite{CT}.

\end{document}